\documentclass[twocolumn,secnumarabic,amssymb, nobibnotes, aps, pra, showkeys]{revtex4-2}
\usepackage[utf8]{inputenc}
\usepackage{amsmath, mathtools, physics, graphicx, siunitx, appendix, verbatim, subcaption, cleveref}

\graphicspath{{figures}}

\newcommand{\Hi}{\tilde{H}_{\text{int}}}
\newcommand{\PMA}{Division of Physics, Math and Astronomy, California Institute of Technology, Pasadena, California 91125, USA}

\begin{document}

\title{The semi-classical Floquet-Markov master equation for Monte Carlo spin integration}
\author{Raymond Tat}
\affiliation{\PMA}

\begin{abstract}
In designing an experiment to measure a neutron electric dipole moment (nEDM), it is often necessary to determine the behavior of an ensemble of spins under time-dependent and randomly fluctuating magnetic fields. This is particularly relevant for the proposed nEDM@SNS experiment, which features ultra-cold neutrons (UCNs) and helium-3 atoms occupying the same measurement cell, as well as a sinusoidal dressing field. In this work, we investigate a new technique to calculate the frequency shifts arising from magnetic field inhomogeneities and motional magnetic fields, particularly in the case where the spins in question are subjected to a time-periodic magnetic field. The method is based on Floquet theory, a general framework for analyzing periodic linear differential equations, and Redfield theory, which governs the time evolution of the density matrix in the presence of weak couplings to an environment. We benchmark the results against the analytical results derived for the static magnetic field case, as well as against the results of a conventional Runge-Kutta integrator, and find agreement with both. We further study the performance of the method, and find order-of-magnitude improvements in runtime over the conventional integrator.
\end{abstract}

\maketitle

\section{Introduction}
The search for physics beyond the Standard Model on the precision frontier necessitates an equally precise understanding of the uncertainties and systematics of the measurement. For an experiment to measure a neutron electric dipole moment (nEDM) for example, detailed knowledge of the systematic frequency shifts associated with the electric field and magnetic field gradients is required~\cite{GolubLamoreauxOverview}. This is facilitated through extensive use of Monte Carlo simulations to investigate the impact of these effects on the experiment; for an nEDM experiment, this may be done by simulating the Bloch equations for a large ensemble neutron spins and observing the frequency shift in simulation. The standard approach for this is to use a numerical integration algorithm such as Runge-Kutta~\cite{Hairer1993} to integrate the Bloch equations for each simulated neutron. Computationally, this is challenging; as the precision increases, so too does the number of simulated particles needed to constrain the statistical uncertainty to the level of the experiment. This difficulty is compounded when time-dependent (dressing) magnetic fields are applied, which have the effect of drastically reducing the step size taken by numerical integrators. Here we propose an alternative approach, which may provide significant gains in performance in cases where the effects of magnetic gradients and electric fields are relatively small. The strategy is to use the Floquet-Markov master equation (FMME)~\cite{Blumel}, an equation which predicts the time evolution of a system subject to a time-periodic Hamiltonian perturbed by a small time-dependent noise term with known power spectrum. Advances in GPU hardware allow this power spectrum to be estimated efficiently with the Monte Carlo method.

\section{Spin Physics and Frequency Shifts}
Nuclear magnetic resonance (NMR)-based EDM experiments require at least two fields: a static (time-independent) magnetic field, which causes the nuclear spin to precess, and a strong electric field which modifies the precession frequency of the nuclei when coupled to an EDM. There may additionally be a time-dependent spin dressing magnetic field as was proposed for the nEDM experiment at the Spallation Neutron Source (nEDM@SNS), which can be used to improve sensitivity \cite{Ahmed_2019}. Because an EDM measurement is a measurement of frequency, it is necessary to exclude spurious sources of frequency shifts. These can arise due to magnetic field gradients, as well as from velocity-dependent effects due to the presence of the electric field. These shifts can be broadly divided into three categories - shifts which are quadratic in gradient magnitude, shifts which are quadratic in electric field strength, and shifts which are linear in gradient and electric field strength. The last shift, often referred to as the ``geometric phase'' or as the ``vxE effect'', poses a great challenge for EDM measurement, as being linear in electric field strength, it can mimic a real EDM. This was first noted in~\cite{PendleburyGeoPhase} and its effect characterized for several import trap geometries in~\cite{GolubLamoreauxGeoPhase}. These shifts have since been studied extensively, for example in~\cite{McGregor},~\cite{Pignol},~\cite{SwankRandomWalks}, or~\cite{Tat}. However, several common scenarios still elude theoretical analysis - for example, no analytic expression is known for the frequency shift for the case where the spins undergo diffuse reflections from the walls of their cell, nor is there a solution incorporating analytically the effects of gravity, which are relevant for ultra-cold neutrons (UCNs). The crux of the difficulty here is in determining the power spectrum of the fluctuating magnetic field, which depends on the correlation functions of the particle trajectories. These correlation functions can only be determined analytically in certain special cases.

\section{Floquet Theory and the Floquet-Markov Master Equation}
Floquet theory describes the time evolution of a quantum system under a time-periodic Hamiltonian in terms of quasi-energies and quasi-eigenstates, which are the time-periodic analogues of the energies and eigenstates of a time-independent Hamiltonian. Specifically, given a Hamiltonian for which $H(t) = H(t + T)$, Floquet's theorem~\cite{FloquetOriginal, GrifoniHanggi} states that solutions to Schr\"odinger's equation have the form
\begin{align}
\label{eq:floquet-states}
  \Psi(t) &= \sum_{\alpha}c_\alpha\Psi_\alpha(t) \\
       &= \sum_{\alpha}c_\alpha\exp(-i\epsilon_\alpha t/\hbar)\Phi_\alpha(t),
\end{align}
where $\Psi_\alpha(t)$ are the Floquet states, and $\Phi_\alpha(t)$ are the Floquet modes, which are time-periodic, i.e. $\Phi_\alpha(t) = \Phi_{\alpha}(t + T)$. $\epsilon_{\alpha}$ are the quasi-energies. Note that the quasi-energies are only defined up to multiples of $2\pi\hbar/T$. The Floquet modes can be computed numerically in several ways, but here we take the approach used in the QuTIP Python Library~\cite{QuTIP}, which is to diagonalize the one-period propagator $U(T)$. The Floquet modes evaluated at $t=0$ are the eigenstates of the one-period propagator. If one now adds a noise term to the Hamiltonian, the time evolution will no longer be given by equation~\ref{eq:floquet-states}, but by working within the basis of Floquet states, one can derive a master equation that describes the non-unitary evolution of the density matrix to second order in the size of the noise term. After applying the Born-Markov and rotating wave approximations as in Redfield theory~\cite{RedfieldTheory, SlichterNMR} one arrives at the Floquet-Markov master equation (FMME)~\cite{Blumel}, which has the form,
\begin{equation}
\label{eq:21}
\dv{t}\widetilde{\rho_{ab}} = R_{abcd}\widetilde{\rho_{cd}},
\end{equation}
where $\widetilde{\rho_{ab}}$ is the density matrix in the basis of the Floquet modes, and $R_{abcd}$ is a time-independent tensor whose elements depend on the power spectrum $S(\omega)$ of the noise term and the Floquet diagonalization of the unperturbed Hamiltonian. In particular, the relevant frequencies for calculating $R_{abcd}$ are of the form $\Delta_{\alpha \beta k} = (\epsilon_{\alpha}- \epsilon_{\beta})/\hbar + k \omega$, with $k$ an integer and $\omega = 2\pi/T$. The set of relevant frequencies depends only on the unperturbed Hamiltonian, while the power spectrum depends only on the noise term. Thus, the FMME allows us to split the task of simulating an ensemble of spins into two parts - first to compute the Floquet diagonalization, and second to estimate the power spectrum of the noise term.

\section{Floquet Diagonalization}
The full derivation of the FMME formalism for this case is described in the appendix. Here we describe the Floquet diagonalization, wherein we find the Floquet modes and quasi-energies, and with these determine the Fourier transforms of the perturbation operators in the Floquet basis. Note that this need only be done once for a given dressing pulse, regardless of how many particles are simulated. As stated previously, the approach to finding the Floquet modes follows the method used by QuTIP~\cite{QuTIP}. This algorithm uses the fact that the Floquet modes obey the relation
\begin{equation}
\label{eq:30}
U(T) \Phi_{\alpha}(0) = \exp(-i\epsilon_{\alpha}T/\hbar)\Phi_{\alpha}(0)
\end{equation}
where $U(T)$ is the one-period propagator (assuming the period is $T$), $\Phi_{\alpha}(0)$ is a Floquet state at $t=0$, and $\epsilon_{\alpha}$ is the corresponding quasi-energy. To find the Floquet states, we begin by solving numerically the Schr\"odinger equation
\begin{equation}
\label{eq:31}
i \hbar \frac{dU(t)}{dt} = H(t)U(t)
\end{equation}
with initial condition $U(0) = 1$ to find $U(T)$. We then diagonalize $U(T)$, taking the eigenvectors to be $\Phi_{\alpha}(0)$ and the phase of the eigenvalues to determine $\epsilon_{\alpha}$. From there we compute the tensor $X_{\alpha \beta k}$, the Fourier transform of the perturbing operators, defined by
\begin{equation}
\label{eq:32}
\bra{\Phi_{\alpha}(t)} \sigma \ket{\Phi_{\beta}(t)} = \sum_{k=-\infty}^{\infty}e^{ik\omega t}X_{\alpha \beta k},
\end{equation}
where $\sigma$ is the perturbing operator (for example, $\sigma$ is the Pauli spin matrix $\sigma_x$ for a noise magnetic field in the $x$ direction). We can invert this relation to find $X_{\alpha \beta k}$,
\begin{equation}
\label{eq:Xexplicit}
X_{\alpha \beta k} = \int_{0}^{T}e^{-ik\omega t}  \bra{\Phi_{\alpha}(t)} \sigma \ket{\Phi_{\beta}(t)} dt/T
\end{equation}
To evaluate this expression, we must compute the Floquet modes evaluated at $t$, rather than zero. The time evolution of the Floquet modes over the first period is given by
\begin{equation}
\label{eq:34}
\Phi_{\alpha}(t) = e^{i\epsilon_{\alpha}t/\hbar}U(t) \Phi_{\alpha}(0),
\end{equation}
where $U(t)$ is found by again integrating the Schr\"odinger equation numerically. From this, we can evaluate the integral in equation~\ref{eq:Xexplicit} using a quadrature rule. Note that the Floquet diagonalization must be computed before the spectrum estimation step, as the relevant frequencies are determined by the quasi-energies; in particular, the relevant frequencies are of the form
\begin{equation}
\label{eq:35}
\Delta_{\alpha \beta k} = (\epsilon_{\alpha} - \epsilon_{\beta})/\hbar + k \omega
\end{equation}

Lastly, as the index $k$ spans all integers, for the sake of computation a finite range of $k$ must be chosen. For this work we assume that the range $-5 \leq k \leq 5$ is sufficient, as according to the analysis in~\cite{Tat}, the contribution to frequency shifts and relaxation for higher values of $k$ (higher frequencies) are suppressed by increasing orders of Bessel functions. It is beneficial to choose as small a range of $k$ as possible, as increasing the range of $k$ increases the computational complexity of the method. However, depending on the application other ranges may be chosen, for example if noise is known to be concentrated around a specific frequency, or if the Larmor and dressing frequency differ greatly from the parameters used here.

\section{Spectrum Estimation}
The standard technique for estimating the power spectrum from evenly-spaced samples is to use the periodogram, which is the magnitude-squared of the discrete Fourier transform (DFT). In order to reduce variance of the estimate, rather than taking the periodogram of one long time series, the periodograms of many shorter windows of the time series are averaged. This is known as Bartlett's method~\cite{Bartlett1948}. As the Floquet-Markov master equation only requires the power spectrum to be known at specific discrete frequencies, we can tune the parameters of the periodogram so that the output frequencies match those needed by the Floquet-Markov master equation. We investigate two algorithms to compute the averaged periodograms - the first is the Goertzel algorithm~\cite{Goertzel}, which is a two-stage filter that computes the DFT at a given frequency. By operating several of these filters in parallel, one can compute as many of the relevant frequencies as is desired. The second approach is to use the fast Fourier transform (FFT). In this case, we simulate an ensemble of particles and collect samples of the noise magnetic field at evenly spaced points in time. We then, at regular intervals, take the FFT magnitude squared of the collected samples. Notably, for both approaches the minimum sampling rate is determined by the Nyquist frequency, while the minimum window length is determined by the desired frequency resolution.

\subsection{Complex Spectrum}
One subtlety involved in this approach to spin integration is that we require not the ordinary power spectrum, defined as the Fourier transform of the correlation function but rather the complex power spectrum, defined as the one-sided Fourier integral of the correlation function. In other words, whereas normally one works with
\begin{equation}
\label{eq:fullSpectrumDefinition}
S(\omega) = \int_{-\infty}^{\infty}\expval{H(t) H(t + \tau)} e^{-i\omega \tau}d\tau
\end{equation}
the FMME instead requires
\begin{equation}
\label{eq:imaginarySpectrumDefinition}
J(\omega) = \int_{0}^{\infty}\expval{H(t) H(t + \tau)} e^{-i\omega \tau}d\tau.
\end{equation}
Of note is the fact that although $S(\omega)$ is purely real, $J(\omega)$ may in general be complex. Without this modification, the FMME would not predict frequency shifts. Because the complex spectrum is needed, both the Goertzel algorithm and FFT approaches will need to be modified in order to estimate $J(\omega)$ instead of $S(\omega)$.

\subsection{Goertzel Algorithm}
\label{subsec:label}
The Goertzel algorithm computes the DFT at a given frequency $\omega$ in two stages. Given samples $x_n$ spaced out with time interval $\Delta t$, the first stage computes
\begin{equation}
\label{eq:goertzel1}
s_n = x_n + 2 \cos(\omega\Delta t) s_{n-1} - s_{n-2}
\end{equation}
and the second stage is
\begin{equation}
\label{eq:goertzel2}
y_n = s_n - e^{-i\omega \Delta t} s_{n-1}
\end{equation}
It can be shown that $y_n$ is related to the DFT at frequency $\omega$, namely
\begin{equation}
\label{eq:25}
y_n = e^{i \omega n \Delta t}\sum_{k = 0}^{n}x_ke^{-i\omega \Delta t k}
\end{equation}

Taking the magnitude squared of $y_n$ results in an estimate of the ordinary power spectrum, which is given by the double summation
\begin{equation}
\label{eq:Ssum}
S(\omega) \approx \frac{\Delta t}{n} \abs{y_n}^2 = \frac{\Delta t}{n} \sum_{j = 0}^{n} \sum_{k = 0}^{n} x_jx_k^{*} e^{-i \omega \Delta t (j - k)},
\end{equation}
whereas to obtain the complex spectrum we need to insert a Heaviside step function into the double sum:
\begin{equation}
\label{eq:Jsum}
J(\omega) \approx \frac{\Delta t}{n}  \sum_{j = 0}^{n} \sum_{k = 0}^{n} x_jx_k^{*} e^{-i \omega \Delta t (j - k)} \Theta( j - k ),
\end{equation}
where
\[
  \Theta(x) = \begin{dcases}
           \label{eq:heavisideCases}
         1 & x > 0 \\
         1/2 & x = 0 \\
         0 & x < 0
       \end{dcases}
\]

Fortunately, the double sum in equation~\ref{eq:Ssum} can be computed efficiently by maintaining the partial sum $y_{m}$ in memory for each $m$. More concretely, we can re-write the double-sum in equation~\ref{eq:Jsum}. Defining $a_{j} = x_{j}e^{-i\omega \Delta t j}$, we get
\begin{align}
  \label{eq:JsumRewritten}
  J(\omega) &\approx \frac{\Delta t}{n} \sum_{j=0}^{n}a_j\sum_{k=0}^{n}a_{k}^{*}\theta(j-k) \\
       &\approx \frac{\Delta t}{n} \sum_{j=0}^{n}a_j\left(-a_j^{*}/2 + \sum_{k=0}^{j}a_k^{*}\right)  \\
       &\approx \frac{\Delta t}{n} \sum_{j=0}^{n}a_j\left(-a_j^{*}/2 + e^{i\omega\Delta t j}y_{j}^{*}\right) \\
       &\approx \frac{\Delta t}{n} \sum_{j=0}^{n}x_j y_j^{*} - x_j^2/2.
\end{align}
From this we can define the partial sum of $J(\omega)$ as $J_m(\omega)$, whose recurrence is given by
\begin{align}
  \label{eq:DoubleSumUnravelled}
  J_{m}(\omega) &= J_{m-1}(\omega) + y_m x_m - \frac{x_{m}^2}{2} \\
  J(\omega) &= \frac{\Delta t}{n} J_n^{*}(\omega).
\end{align}
Combining the recurrence relations of equations~\ref{eq:goertzel1},~\ref{eq:goertzel2}, and~\ref{eq:DoubleSumUnravelled}, we arrive at an algorithm which computes the imaginary spectrum $J(\omega)$ which is linear in time and constant in memory usage. This process must be repeated for each frequency $\omega$, and for each generated time series.


\subsection{Fast Fourier Transform (FFT)}
Alternatively, one can collect samples of the noise in an array, then take a batched FFT of the array. Algorithmically, this corresponds to
\begin{equation}
  \label{}
  S(\omega) = \abs{\operatorname{FT}[\Vec{x}]}^2,
\end{equation}
where $\Vec{x}$ is the vector ${x_0, x_1 \ldots x_n}$, and $\operatorname{FT}$ is the Fourier transform. This is repeated and for each generated time series and the results averaged. Generally, the frequencies output by the Fourier transform will not correspond exactly to those needed by the FMME. Thus, some interpolation is necessary. For simplicity, we use nearest-neighbor interpolation.
In order to find the complex spectrum, we can use the fact that the spectrum is the Fourier transform of the correlation function. Thus, we can take the inverse Fourier transform, multiply by the Heaviside step function in the time domain, then Fourier transform back to the frequency domain. In other words, we compute
\begin{align}
  \label{}
  R(\tau) &= \operatorname{FT}^{-1}[S(\omega)] \\
  J(\omega) &= \operatorname{FT}[\Theta(\tau)R(\tau)].
\end{align}
These last two steps are performed only once per ensemble, after averaging to obtain an estimate of $S(\omega)$.

\section{Acceleration with Graphics Processing Units (GPUs)}
Simulating enough particles to constrain frequency shifts to the level of precision needed for experiments such as nEDM@SNS requires on the order of millions of particles to be simulated for a thousand seconds each. In the field of UCN, a variety of open source code packages have been developed which can integrate the Bloch equations along a particle's trajectory: for example starUCN~\cite{STARucn}, MCUCN~\cite{MCUCN}, PENTrack~\cite{PENTrack}, Geant4~\cite{AGOSTINELLI2003250}, and Kassiopeia~\cite{Kassiopeia}. While these software packages have a variety of advantages in terms of simulating arbitrary geometries, material interactions, or sophisticated wall reflection models, for the specific purpose of simulating large numbers of particles, generating particle trajectories serially will be too time-consuming. To expedite this process, we leverage general-purpose graphical processing unit (GPGPU) programming. Each GPU thread will be responsible for tracking the trajectory of a single particle, sampling the magnetic field at even time intervals along its path. This allows many particles to be simulated in parallel, reducing the computation time by orders of magnitude at the cost of generality of the simulation software. For the FFT approach, using GPUs offers an additional advantage, as in this case we can use existing CUDA libraries to efficiently perform the requisite FFT and tensor reductions on NVIDIA GPUs. These libraries are cuFFT, which performs FFT; and cuTENSOR, which performs general tensor contractions and reductions. For GPGPU programming, some additional optimizations may be necessary to extract maximum performance out of GPU hardware; specifically, we ensure that the tensor which contains the noise samples is arranged so that samples collected consecutively in time are adjacent in memory. Further optimizations were required for the GPU-based particle trajectory tracking scheme, which are detailed in~\cite{mjmThesis}.

\section{Comparison with Theory}
\begin{figure}[!htbp]
  \centering
  \includegraphics[width=0.9\linewidth]{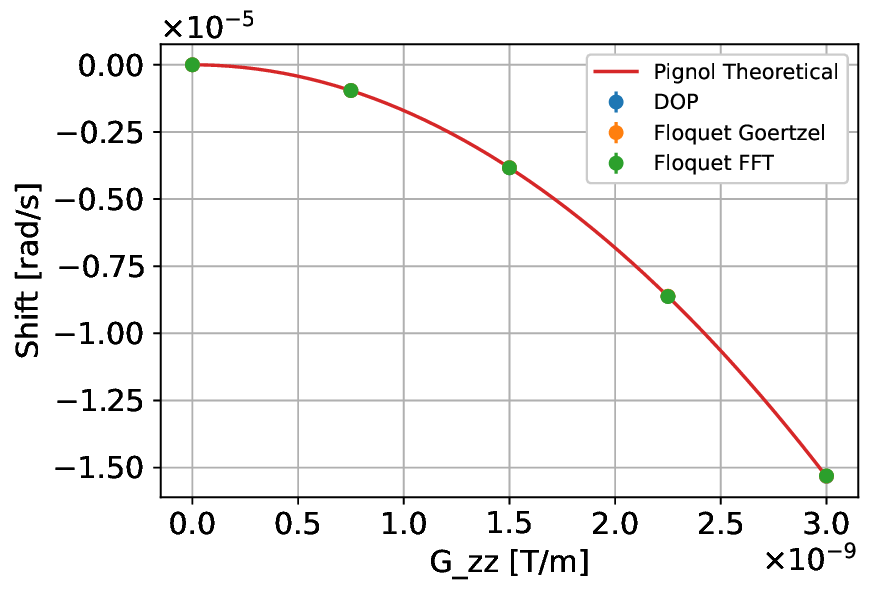}
  \caption{Comparison of DOP, FMME Goertzel, and FMME FFT methods with theoretical predictions for the $B^2$ frequency shift in the adiabatic limit.}
  \label{fig:B2}
\end{figure}

To confirm that the output of the FMME-based algorithms agree with the theoretical frequency shifts, we compare with the analytical results found in~\cite{Pignol} in the adiabatic limit for free precession. We also benchmark these simulations against a standard Runge-Kutta algorithm, Dormand-Prince 853 (DOP)~\cite{Hairer1993}. We show comparisons for $B^2$ (figure~\ref{fig:B2}), $E^2$ (figure~\ref{fig:E2}), and linear-in-E (figure~\ref{fig:EB}) frequency shifts. Table~\ref{table:params} in Appendix~\ref{appendix:params} lists the parameters used for these simulations.

\begin{figure}[!htbp]
  \centering
  \includegraphics[width=0.9\linewidth]{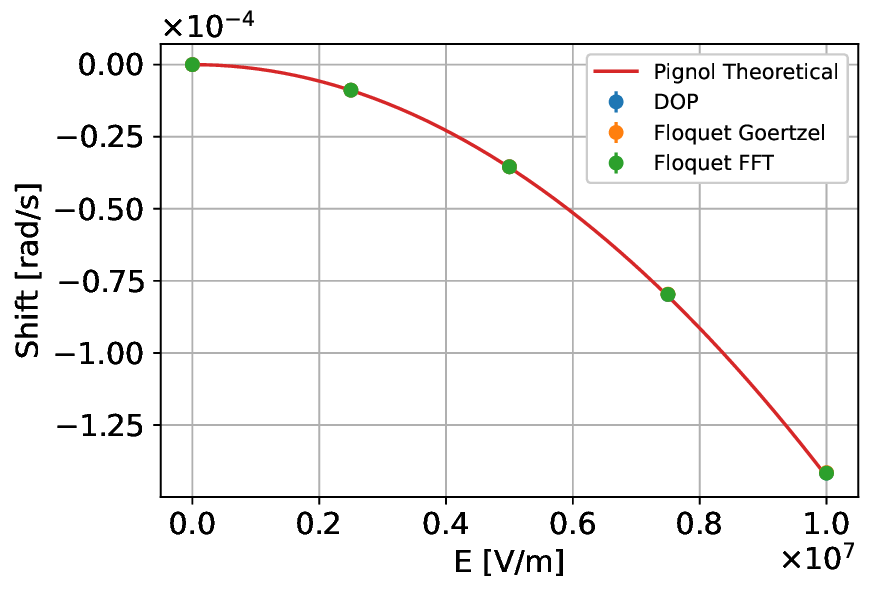}
  \caption{Comparison of DOP, FMME Goertzel, and FMME FFT methods with theoretical predictions for the $E^2$ frequency shift in the adiabatic limit.}
  \label{fig:E2} 
\end{figure}

\begin{figure}[!htbp]
  \centering
  \includegraphics[width=0.9\linewidth]{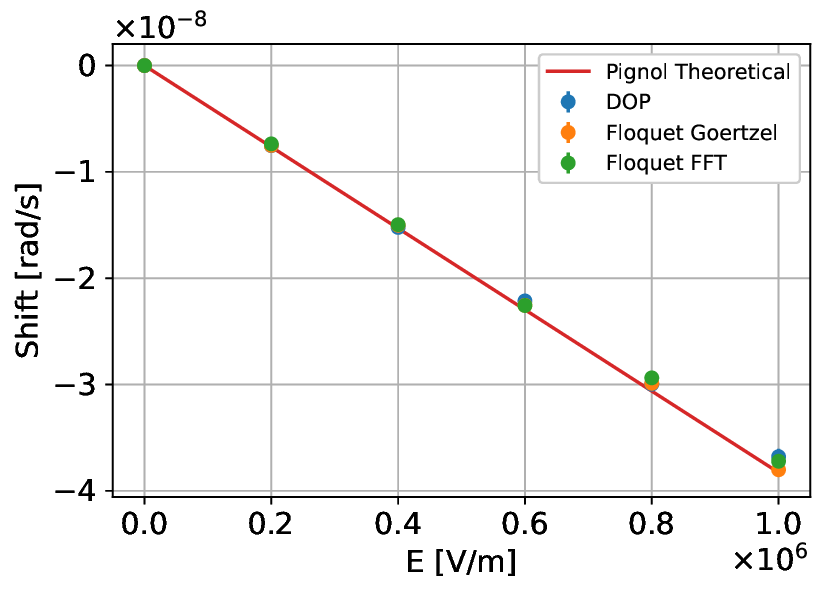}
  \caption{Comparison of DOP, FMME Goertzel, and FMME FFT methods with theoretical predictions for the linear-in-E frequency shift in the adiabatic limit. For these simulations the electric field is varied, while the gradient is fixed at $\frac{dB_{z}}{dz} = \SI{3e-10}{\tesla\per\meter}$}
  \label{fig:EB}
\end{figure}

\begin{figure}[!htbp]
  \centering
  \includegraphics[width=0.9\linewidth]{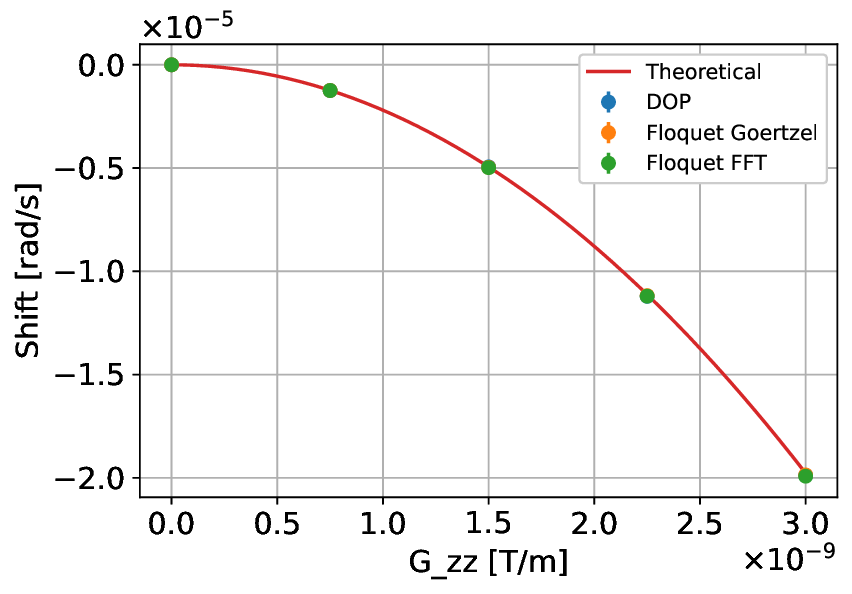}
  \caption{Comparison of DOP, FMME Goertzel, and FMME FFT methods with theoretical predictions for the $B^2$ frequency shift in the adiabatic limit for spin dressing.}
  \label{fig:B2Dress}   
\end{figure}

Likewise, in figure~\ref{fig:B2Dress} we compare the output of the FMME-based algorithms with the results given by the theory in~\cite{Tat} for the spin dressing case. As the authors of~\cite{Tat} consider only perturbations along the dressing field direction, only the $B^2$ shift can be treated theoretically. 

\subsection{Effect of Window Length}
\begin{figure}[!htbp]
  \centering
  \includegraphics[width=0.9\linewidth]{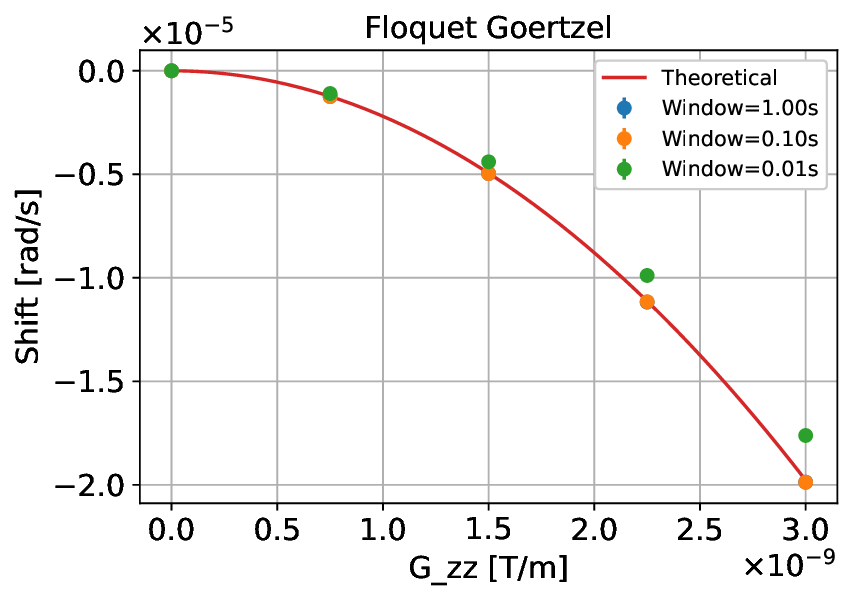}
  \caption{The $B^2$ shift estimated by the Floquet Goertzel method varies with window length. }
  \label{fig:B2windowGoertzel}
\end{figure}

\begin{figure}[!htbp]
  \centering
  \includegraphics[width=0.9\linewidth]{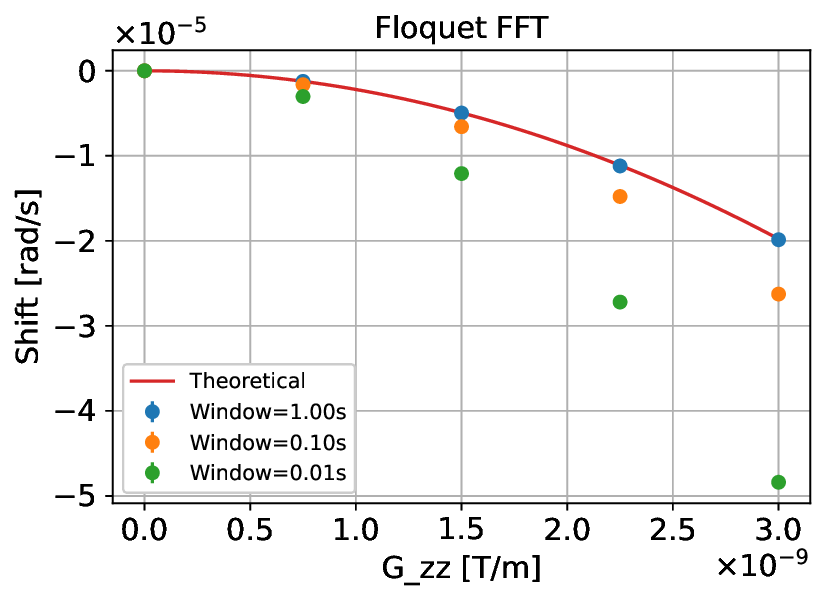}
  \caption{The $B^2$ shift estimated by the Floquet FFT method varies with window length. }
  \label{fig:B2windowFFT}
\end{figure}

In cases where the characteristic correlation time of the system is sufficiently long, the window length must be increased to fully capture the correlation function. This is particularly import for estimating the $B^2$ shift, as in this case the relevant correlation time is the time taken for the particles to diffuse across the cell, which for the parameters used for these simulations is on the order of one second for helium-3 atoms (this time strongly depends on temperature). We illustrate this phenomenon in figures~\ref{fig:B2windowGoertzel} and~\ref{fig:B2windowFFT}, where we can see that decreasing the window length has the effect of increasing the bias for the Floquet-based methods. The bias is suppressed as the window length approaches the typical correlation time of the system.

\section{Performance}
\begin{figure}[!htbp]
  \centering
  \includegraphics[width=\linewidth]{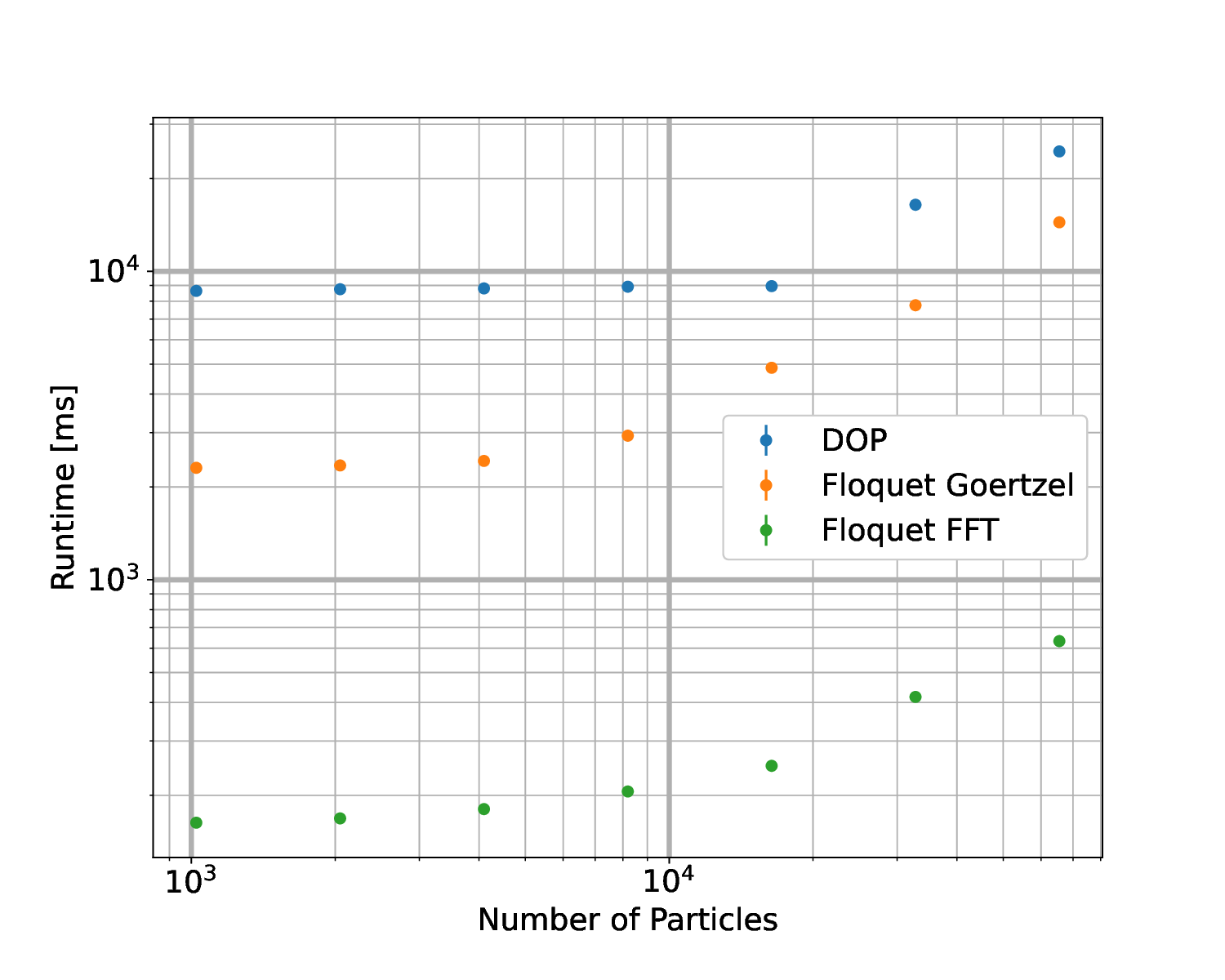}
  \caption{ Comparison of time taken to simulate various numbers of particles with DOP853, FMME Goertzel method, or FMME FFT method.}
  \label{fig:performance}
\end{figure}

Figure~\ref{fig:performance} compares the performance of the Goertzel algorithm and FFT approach with that of an adaptive Runge-Kutta integrator, DOP853. The parameters used for performance testing are identical to those used for the theory comparison, except that the dressing amplitude is \SI{3.0e-5}{\tesla}, the dressing frequency is \SI{1}{\kilo\hertz}, and the window length is shortened to 0.1 seconds. Note that whereas the Runge-Kutta integrator outputs the entire ensemble of spins, the FMME-based algorithms output only the density matrix. For similar number of particles, the Goertzel algorithm-based method outperforms the Runge-Kutta based approach, while the FFT-based approach outperforms both.

\section{Conclusion}
In this work, we developed a new method to integrate the Bloch equation for large ensembles of spin-1/2 particles in a time-periodic magnetic field subject to noise. Our approach is to use the Floquet-Markov Master Equation to divide the problem of integrating a large number of differential equations into two subproblems: first, to compute the Floquet states and quasi-energies of the (noise-free) periodic Hamiltonian; and second, to compute the complex power spectrum of the noise. The former is computationally trivial, while the second can be computed efficiently using fast Fourier transform algorithms on modern hardware. To this end, we also adapt the Floquet-Markov master equation to the semi-classical case, in which the environment noise is described by a classical correlation function rather than as a quantum mechanical bath. While this algorithm is appliable in general to any quantum system where both a periodic driving field and random fluctuations are present, it is particularly well-suited to simulating the frequency shifts and relaxations which are relevant to precision measurements experiments such as nEDM@SNS. We compare the results of the method when computing these frequency shifts both to a standard Runge-Kutta integrator and theoretical results based on the work of~\cite{Pignol}, and find good agreement with both. We additionally compare the runtime of the various algorithms, and find an order-of-magnitude improvement over the standard Runge--Kutta algorithm when using the FFT-based Floquet method, while the runtime savings with the Goertzel's algorithm-based Floquet method are more modest.

\section{Acknowledgments}
The author would like to thank Brad Filippone, Robert Golub, David Davis, and the nEDM@SNS publications committee for their helpful input and review of the paper. The author would also like to thank Matthew Morano and David Mathews for their work in designing the GPU-based particle tracking software on which this work is based. This work was funded by the National Science Foundation (NSF) grants 2110898 and 1822515. This research was supported in part through research cyberinfrastructure resources and services provided by the Partnership for an Advanced Computing Environment (PACE) at the Georgia Institute of Technology, Atlanta, Georgia, USA. This research was supported in part through research infrastructure and services provided by the Rogues Gallery testbed hosted by the Center for Research into Novel Computing Hierarchies (CRNCH) at Georgia Tech. The Rogues Gallery testbed is primarily supported by the National Science Foundation (NSF) under NSF Award Number \#2016701. Any opinions, findings and conclusions, or recommendations expressed in this material are those of the author, and do not necessarily reflect those of the NSF.

\onecolumngrid
\appendix
\section{Simulation Parameters}
\label{appendix:params}
Table~\ref{table:params} lists the parameters used for comparisons with the theoretical frequency shifts predicted by~\cite{Pignol}.
\begin{table}[!h]
  \begin{center}
    \begin{tabular}{c | c | c}
      Parameter & Value & Unit \\
      \hline
      B0 & $5.0 \times 10^{-6}$ & \unit{\tesla} \\
      Dressing Amplitude & 0 & \unit{\tesla} \\
      Dressing Frequency & N/A & \unit{\hertz} \\
      Cell Dimensions & $1.0 \times 1.0 \times 1.0$ & \unit{\meter} \\
      Particle Mass & $1.20239 \times 10^{-26}$ & \unit{\kilo\gram} \\
      Simulation Time & $1.0$ & \unit{\second} \\
      Window Length & $1.0$ & \unit{\second} \\
      Gyromagnetic Ratio & $-2.038 \times 10^8$ & \unit{\per\second\per\tesla} \\
      Sampling Rate & $10^4$ & \unit{\hertz} \\
      Temperature & 0.25 & \unit{\kelvin} \\
      Wall diffusivity & 0.0 & \\
      Gravity & 0.0 & \unit{\meter\per\second\squared} \\
    \end{tabular}
  \end{center}
  \caption{Parameters used for comparisons with~\cite{Pignol}.}
  \label{table:params}
\end{table}

\section{Floquet-Markov Master Equation derivation}
In this section, we derive the semi-classical Floquet-Markov master equation including energy shifts. This derivation largely follows that in~\cite{Blumel}.

\subsection{Von Neumann-Liouville Equation in the Floquet Basis}
Consider a Hamiltonian
\begin{equation}
\label{eq:3}
H(t) = H_0(t) + H_{\text{int}}(t),
\end{equation}

where $H_0(t)$ is periodic and deterministic with period $T = 2\pi/\omega$, and $H_{\text{int}}(t)$ is a stochastic component given by
\begin{equation}
\label{eq:2}
H_{\text{int}}(t) = f(t) \sigma
\end{equation}
for some (quantum-mechanical) operator $\sigma$ and some wide-sense stationary (WSS) scalar stochastic process of zero mean $f(t)$ whose autocorrelation is known.

We begin with the von Neumann-Liouville equation in the interaction picture:
\begin{equation}
\label{eq:1}
i\hbar \frac{d\tilde{\rho}}{dt} = [\Hi(t), \tilde{\rho}]
\end{equation}
Where $\tilde{\rho}$ and $\Hi$ are the density matrix and Hamiltonian, each evaluated in the interaction picture:
\begin{align}
  \tilde{\rho}(t) &= U^{\dag}(t) \rho(t) U(t) \\
  \Hi(t) &= U^{\dag}(t) H_{\text{int}}(t) U(t)
\end{align}
The transformation to the interaction picture is given in the Floquet basis:
\begin{align}
U(t) = \sum_{\alpha}e^{-i\mu_{\alpha}t}\ket{\Phi_{\alpha}(t)}\bra{\Phi_{\alpha}(0)}
\end{align}
where the $\ket{\Phi_{\alpha}(t)}$ are the Floquet modes and the $\mu_{\alpha} = \epsilon_{\alpha}/\hbar$ are the frequencies associated with the corresponding quasi-energies. We can expand the interaction Hamiltonian in the Floquet basis, yielding
\begin{align}
  \Hi(t) &= U^{\dag}(t) f(t) \sigma U(t) \\
         &= f(t) \sigma_{\text{int}}(t) \label{eq:fsigma}\\
         &= f(t)\sum_{\alpha, \beta}e^{i(\mu_{\alpha} - \mu_{\beta})t}\ket{\Phi_{\alpha}(0)}\bra{\Phi_{\alpha}(t)} \sigma \ket{\Phi_{\beta}(t)}\bra{\Phi_{\beta}(0)},
\end{align}
where $\sigma_{\text{int}}(t) = U^{\dag}(t) \sigma U(t)$ is $\sigma$ in the interaction picture.

Expanding the von Neumann-Liouville equation, we get:
\begin{align}
  \frac{d\tilde{\rho}}{dt} &= \frac{1}{i\hbar} [\Hi(t), \tilde{\rho}(t)] \\
                        &= \frac{1}{i\hbar} \left[\Hi(t), \tilde{\rho}(0) + \frac{1}{i\hbar}\int_0^tdt'\left[\Hi(t'), \tilde{\rho}(t')\right]\right] \\
                        &= \frac{1}{i\hbar} \left[\Hi(t), \tilde{\rho}(0)\right] + \frac{1}{(i\hbar)^2} \int_0^tdt' \left[\Hi(t), \left[\Hi(t'), \tilde{\rho}(t')\right]\right].
\end{align}
Substituting the expression in~\ref{eq:fsigma}, this becomes
\begin{align}
  \frac{d\tilde{\rho}}{dt} &= \frac{1}{i\hbar} f(t) \left[\sigma_{\text{int}}(t), \tilde{\rho}(0)\right] + \frac{1}{(i\hbar)^2} \int_0^tdt' f(t) f(t') \left[\sigma_{\text{int}}(t), \left[\sigma_{\text{int}}(t'), \tilde{\rho}(t')\right]\right]. \\
  &=  \frac{1}{(i\hbar)^2} \int_0^tdt' R(t - t') \left[\sigma_{\text{int}}(t), \left[\sigma_{\text{int}}(t'), \tilde{\rho}(t)\right]\right].
\end{align}
where in the second line we take the average over the functions $f(t)$ $(\expval{f(t)} == 0, \expval{f(t)f(t')} = R(t - t'))$ and apply the Markov approximation, replacing $\tilde{\rho}(t')$ with $\tilde{\rho}(t)$. Next we introduce the Fourier transform $X_{\alpha \beta k}$ of the matrix element $\bra{\Phi_{\alpha}(t)} \sigma \ket{\Phi_{\alpha}(t)}$, defined by
\begin{equation}
\label{eq:5}
\bra{\Phi_{\alpha}(t)} \sigma \ket{\Phi_{\beta}(t)} = \sum_{k=-\infty}^{\infty}e^{ik\omega t}X_{\alpha \beta k}.
\end{equation}
We can now write $\sigma_{\text{int}}(t)$ as
\begin{align}
\label{eq:6}
\sigma_{\text{int}}(t) &= \sum_{\alpha, \beta, k}e^{i(\mu_{\alpha}-\mu_{\beta} + k\omega)t}\ket{\Phi_{\alpha}(0)}\bra{\Phi_{\beta}(0)}X_{\alpha \beta k} \\
&= \sum_{\alpha, \beta, k}e^{i\Delta_{\alpha \beta k}t}\ket{\Phi_{\alpha}(0)}\bra{\Phi_{\beta}(0)}X_{\alpha \beta k},
\end{align}
where we have defined $\Delta_{\alpha \beta k} = \mu_{\alpha}-\mu_{\beta} + k\omega$. Substituting into the von Neumann-Liouville equation we obtain
\begin{multline}
\frac{d\tilde{\rho}}{dt} = \frac{1}{(i\hbar)^2} \int_0^tdt' R(t - t')\left[ \sum_{\alpha, \beta, k}e^{i\Delta_{\alpha \beta k}t}\ket{\Phi_{\alpha}(0)}\bra{\Phi_{\beta}(0)}X_{\alpha \beta k}, \right. \\ \left. \left[ \sum_{\alpha', \beta', k'}e^{i\Delta_{\alpha' \beta' k'}t'}\ket{\Phi_{\alpha'}(0)}\bra{\Phi_{\beta'}(0)}X_{\alpha' \beta' k'}, \tilde{\rho}(t) \right] \right].
\end{multline}
Rearranging terms, we get
\begin{multline}
\frac{d\tilde{\rho}}{dt} = \frac{1}{(i\hbar)^2} \int_0^tdt' R(t - t') \sum_{\alpha, \beta, k} \sum_{\alpha', \beta', k'}e^{i\Delta_{\alpha \beta k}t + i\Delta_{\alpha' \beta' k'}t'}X_{\alpha \beta k} X_{\alpha' \beta' k'} \\
\left[\ket{\Phi_{\alpha}(0)}\bra{\Phi_{\beta}(0)}, \left[ \ket{\Phi_{\alpha'}(0)}\bra{\Phi_{\beta'}(0)}, \tilde{\rho}(t) \right] \right]
\end{multline}
We now make use of some useful symmetries of $\Delta$ and $X$:
\begin{align}
  \label{eq:symmetries}
  \Delta_{\alpha \beta k} &= - \Delta_{\beta \alpha -k} \\
  X_{\alpha \beta k} &= X^{*}_{\beta \alpha -k}.
\end{align}
Further note that since $k'$ ranges from $-\infty$ to $+\infty$, we are free to replace $k'$ with $-k'$ in the sum. And since $\alpha'$ and $\beta'$ each index over the same set (the Floquet states), we can freely exxchange $\alpha'$ and $\beta'$. Applying all of these changes, we get:

\begin{align}
  \frac{d\tilde{\rho}}{dt} &= \begin{multlined}[t]
         \frac{1}{(i\hbar)^2} \int_0^tdt' R(t-t') \sum_{\alpha, \beta, k} \sum_{\alpha', \beta', k'}e^{i\Delta_{\alpha \beta k}t - i \Delta_{\alpha' \beta' k'}t'}X_{\alpha \beta k} X_{\alpha' \beta' k'}^{*} \\
        \left[\ket{\Phi_{\alpha}(0)}\bra{\Phi_{\beta}(0)}, \left[ \ket{\Phi_{\beta'}(0)}\bra{\Phi_{\alpha'}(0)}, \tilde{\rho}(t) \right] \right]
      \end{multlined} \\
                        &= \begin{multlined}[t]
         \frac{1}{(i\hbar)^2} \sum_{\alpha, \beta, k} \sum_{\alpha', \beta', k'} e^{i(\Delta_{\alpha, \beta, k} - \Delta_{\alpha' \beta' k'})t} \left(\int_0^tdt' R(t-t') e^{i\Delta_{\alpha' \beta' k'}(t-t')}\right) \\ 
        X_{\alpha \beta k} X_{\alpha' \beta' k'}^{*} \left[\ket{\Phi_{\alpha}(0)}\bra{\Phi_{\beta}(0)}, \left[ \ket{\Phi_{\beta'}(0)}\bra{\Phi_{\alpha'}(0)}, \tilde{\rho}(t) \right] \right]
      \end{multlined}
\end{align}
We further apply the Markovian approximation, assuming that correlation times are short compared to the measurement time. For particles confined to a container, the typical timescales involved are the average time between collisions and the diffusion time across a container. For example for the nominal parameters of the nEDM experiment in~\cite{Ahmed_2019}, the average collision time is on the order of milliseconds, while the diffusion time is on the order of a second - both of which are much shorter than the proposed measurement time of \SI{1000}{\second}. In such cases $R(\tau)$ decays quickly enough that we can replace the lower limit of integration with $-\infty$. Thus,
\begin{align}
  \int_0^tdt' R(t-t') e^{i\Delta_{\alpha' \beta' k'}(t-t')} &\approx  \int_{-\infty}^tdt' R(t-t') e^{i\Delta_{\alpha' \beta' k'}(t-t')} \\
                                           &= -\int_{\infty}^0 d\tau R(\tau) e^{i\Delta_{\alpha' \beta' k'}\tau} \\
                                           &= \int_0^{\infty} d\tau R(\tau) e^{i\Delta_{\alpha' \beta' k'}\tau} \\
                                           &= J(\Delta_{\alpha' \beta' k'})
\end{align}
where we have defined $J(\omega) \equiv \int_0^{\infty}d\tau R(\tau) e^{i\omega \tau}$ as the one-sided Fourier transform of the autocorrelation function. At this point, we can observe that $J(\omega)$ is related to what we refer to simply as the spectrum, defined as $S(\omega) \equiv \int_{-\infty}^{\infty}d\tau R(\tau) e^{i\omega \tau}$. Notice the integral goes from $-\infty$ to $+\infty$. Let us note some properties of $J(\omega)$ and $S(\omega)$:

\begin{itemize}
\item $S(\omega)$ is real and even (because it is the Fourier transform of $R(\tau)$, which is real and even)
\item $J(\omega) = J^{*}(-\omega)$
\item $S(\omega) = J(\omega) + J^{*}(\omega) = 2 \Re[J(\omega)]$
\item $J(\omega) = \frac{1}{2}S(\omega) - \frac{i}{2\pi}\mathcal{P}\int_{-\infty}^{\infty}\frac{S(\omega')}{\omega' - \omega}d \omega'$ ($\mathcal{P}$ denotes the Cauchy principal value)
\end{itemize}
The first three identities can be seen directly from the definitions of $J(\omega)$ and $S(\omega)$. The final identity can be derived either from the Kramers-Kronig relations, or by using the Fourier transform of the Heaviside step function (i.e. $J(\omega) = \int_{-\infty}^{\infty}d\tau \Theta (\tau)R(\tau) e^{i\omega \tau}$) along with the convolution theorem.

To simplify notation, we also introduce the operators
\begin{equation}
\label{eq:8}
\Gamma_{\alpha\beta} \equiv \ket{\Phi_{\alpha}(0)}\bra{\Phi_{\beta}(0)}
\end{equation}

With this, our master equation has now simplified to

\begin{multline}
\frac{d\tilde{\rho}}{dt} = \frac{1}{(i\hbar)^2}  \sum_{\alpha, \beta, k} \sum_{\alpha', \beta', k'} J(\Delta_{\alpha' \beta' k'}) e^{i(\Delta_{\alpha, \beta, k} - \Delta_{\alpha' \beta' k'})t} X_{\alpha \beta k} X_{\alpha' \beta' k'}^{*} \\
\left[\Gamma_{\alpha\beta}, \left[ \Gamma_{\beta'\alpha'}, \tilde{\rho}(t) \right] \right]
\end{multline}

Next, we need to apply a rotating wave approximation. The argument is that the phase factor $e^{i(\Delta_{\alpha, \beta, k} - \Delta_{\alpha' \beta' k'})t}$ oscillates quickly, and thus can be neglected unless $\Delta_{\alpha, \beta, k} = \Delta_{\alpha' \beta' k'}$. From the definition of $\Delta_{\alpha \beta k} = \mu_\alpha - \mu_\beta + k\omega$ we can see that there are two cases where these two frequencies are equal:
\begin{enumerate}
\item $(\alpha, \beta, k) = (\alpha', \beta', k')$ \label{case:a}
\item $\alpha = \beta, \alpha' = \beta', k = k'$
\end{enumerate}
Some quantum systems may have additional degeneracies, but we will assume these are the only ones. We tackle case~\ref{case:a} first.
\subsubsection{$(\alpha, \beta, k) = (\alpha', \beta', k')$}
\label{subsec:rwa}
Assuming $(\alpha, \beta, k) = (\alpha', \beta', k')$, the master equation further simplifies to:

\begin{align}
  \frac{d\tilde{\rho}}{dt} &= \frac{1}{(i\hbar)^2}  \sum_{\alpha, \beta, k}  J(\Delta_{\alpha \beta k}) \abs{X_{\alpha \beta k}}^2 \left[\Gamma_{\alpha\beta}, \left[ \Gamma_{\beta\alpha}, \tilde{\rho}(t) \right] \right] \\
                        &= -\sum_{\alpha, \beta} \zeta_{\alpha \beta} \left[\Gamma_{\alpha\beta}, \left[ \Gamma_{\beta\alpha}, \tilde{\rho}(t) \right] \right] \\
                        &= -\sum_{\alpha, \beta} \zeta_{\alpha \beta} \left(\Gamma_{\alpha\alpha}\rho + \rho\Gamma_{\beta\beta} - \Gamma_{\alpha\beta}\rho\Gamma_{\beta\alpha} - \Gamma_{\beta\alpha}\rho\Gamma_{\alpha\beta}\right)
\end{align}
where in the second line we perform the sum over $k$, and define
\begin{equation}
\label{eq:9}
\zeta_{\alpha \beta} \equiv \frac{1}{\hbar^2}  \sum_{k}  J(\Delta_{\alpha \beta k}) \abs{X_{\alpha \beta k}}^2
\end{equation}
Examining $\zeta_{\alpha \beta}$, we see that $\zeta_{\alpha \beta} = \zeta_{\beta \alpha}^{*}$, i.e. $\zeta$ is Hermitian. This matrix is related to the matrix $M_{\alpha\beta}$ given in~\cite{Blumel}. For classical noise, we should take $n_{th}(\omega) = \frac{1}{2\pi\hbar^2} >> 1 $. We obtain:
\begin{align}
\label{eq:13}
  M_{\alpha\beta} &= \frac{1}{\hbar^2}\sum_{\alpha, \beta, k}\Theta(\Delta_{\alpha \beta k})S(\Delta_{\alpha \beta k}) \abs{X_{\alpha \beta k}}^2 + \Theta(\Delta_{\beta \alpha k})S(\Delta_{\beta \alpha k}) \abs{X_{\beta \alpha k}}^2 \\
         &= \frac{1}{\hbar^2}\sum_{\alpha, \beta, k}\Theta(\Delta_{\alpha \beta k})S(\Delta_{\alpha \beta k}) \abs{X_{\alpha \beta k}}^2 + \Theta(-\Delta_{\alpha \beta k})S(\Delta_{\alpha \beta k}) \abs{X_{\alpha \beta k}}^2 \\
         &= \frac{1}{\hbar^2}\sum_{\alpha, \beta, k}S(\Delta_{\alpha \beta k}) \abs{X_{\alpha \beta k}}^2 \\
         &= \zeta_{\alpha, \beta} + \zeta_{\alpha, \beta}^{*} \\
           &= \zeta_{\alpha, \beta} + \zeta_{\beta, \alpha}
\end{align}
Notably, if we assume $J(\omega)$ is purely real, which amounts to discarding the principal part of the integral as they do in~\cite{Blumel}, then $M_{\alpha\beta}$ and $\zeta_{\alpha\beta}$ are proportional. We now examine the matrix elements of $\tilde{\rho}$, which we will define as $\rho_{\nu\sigma} = \bra{\Phi_{\nu}(0)} \tilde{\rho} \ket{\Phi_{\sigma}(0)}$. Taking the matrix elements of both sides, we get
\begin{align}
\label{eq:11}
  \frac{d\rho_{\nu\sigma}}{dt} &= - \sum_{\alpha, \beta} \zeta_{\alpha \beta} \left( \delta_{\nu\alpha}\rho_{a\sigma} + \rho_{\nu\beta} \delta_{\beta\sigma} - \delta_{\nu\alpha}\delta_{\alpha\sigma}\rho_{\beta\beta} - \delta_{\nu\beta}\delta_{\beta\sigma}\rho_{\alpha\alpha}\right) \\
                     &= - \sum_{\beta} \zeta_{\nu\beta} \rho_{\nu\sigma} - \sum_{\alpha}\zeta_{\alpha \sigma}\rho_{\nu\sigma} + \delta_{\nu\sigma} \sum_{\beta}\zeta_{\nu\beta}\rho_{\beta\beta} + \delta_{\nu\sigma} \sum_{\alpha}\zeta_{\alpha\sigma}\rho_{\alpha\alpha} \\
                     &= - \left(\sum_{\alpha} \zeta_{\nu \alpha} + \zeta_{\alpha \sigma}\right) \rho_{\nu\sigma} + \delta_{\nu\sigma} \left(\sum_{\alpha}(\zeta_{\nu\alpha} + \zeta_{\alpha\sigma})\rho_{\alpha\alpha} \right)
\end{align}

We can put this expression into perhaps a more intuitive form if we split this equation into the diagonal and off-diagonal cases: $\nu = \sigma$ and $\nu \neq \sigma$.
\begin{align}
  \frac{d\rho_{\nu\nu}}{dt} &=  \sum_{\alpha}[\zeta + \zeta^T]_{\nu\alpha}\rho_{\alpha\alpha} - \sum_{\alpha}[\zeta + \zeta^T]_{\nu\alpha}\rho_{\nu\nu}\\
  \frac{d\rho_{\nu\sigma}}{dt} &= -\left(\sum_{\alpha} \zeta_{\nu \alpha} + \zeta_{\alpha \sigma}\right) \rho_{\nu\sigma} , \quad \nu \neq \sigma
\end{align}

Finally, to match more closely with the notation in~\cite{Blumel}, we switch the labelling of the indices.

\begin{align}
  \frac{d\rho_{\alpha\alpha}}{dt} &=  \sum_{\nu}[\zeta + \zeta^T]_{\alpha\nu}\rho_{\nu\nu} - \sum_{\nu}[\zeta + \zeta^T]_{\alpha\nu}\rho_{\alpha\alpha}\\
  \frac{d\rho_{\alpha\beta}}{dt} &= -\left(\sum_{\nu} \zeta_{\alpha\nu} + \zeta_{\nu\beta}\right) \rho_{\alpha\beta} , \quad \alpha \neq \beta
\end{align}

\subsubsection{$\alpha = \beta, \alpha' = \beta', k = k', \alpha \neq \alpha'$}
\label{subsec:rwa2}
In this case we have
\begin{equation}
  \frac{d\tilde{\rho}}{dt} = \frac{1}{(i\hbar)^2}  \sum_{\alpha \neq \alpha', k}  J(\Delta_{\alpha' \alpha' k}) X_{\alpha \alpha k} X_{\alpha' \alpha' k}^{*}  \left[\Gamma_{\alpha\alpha}, \left[ \Gamma_{\alpha'\alpha'}, \tilde{\rho}(t) \right] \right]
\end{equation}

It is worth noting that $\Delta_{\alpha \alpha k} = k\omega$ regardless of $\alpha$, and that $X_{\alpha \alpha k} = X_{\alpha \alpha -k}^{*}$. Let us now define the matrix
\begin{align}
\label{eq:12}
\xi_{\alpha\alpha'} &= \frac{1}{\hbar^2}  \sum_{k}  J(\Delta_{\alpha' \alpha' k}) X_{\alpha \alpha k} X_{\alpha' \alpha' k}^{*} \quad \alpha \neq \alpha'\\
        &= \frac{1}{\hbar^2}  \sum_{k}  J(k\omega) X_{\alpha \alpha k} X_{\alpha' \alpha' k}^{*} \quad \alpha \neq \alpha'
\end{align}
Also, since we are summing only over terms with $\alpha \neq \alpha'$, we have that $\Gamma_{\alpha\alpha}\Gamma_{\alpha'\alpha'} = \Gamma_{\alpha'\alpha'}\Gamma_{\alpha\alpha} = 0$. Therefore we get
\begin{align}
\label{eq:14}
  \frac{d\tilde{\rho}}{dt} &=  \sum_{\alpha \neq \alpha'} \xi_{\alpha\alpha'}  \left( \Gamma_{\alpha\alpha} \tilde{\rho} \Gamma_{\alpha'\alpha'} + \Gamma_{\alpha'\alpha'} \tilde{\rho} \Gamma_{\alpha\alpha} \right) \\
  \frac{d\rho_{\alpha\beta}}{dt} &=  \xi_{\alpha\beta}\rho_{\alpha\beta} + \xi_{\beta\alpha}\rho_{\beta\alpha} \quad \alpha \neq \beta
\end{align}

\subsection{Fields in Multiple Directions}
Up until now we have considered only perturbations which apply to a single operator. What if there are multiple operators involved (e.g. in the case of spin, multiple magnetic fields)? Then there may be correlations between the perturbations for the various operators.

Suppose now that our perturbation Hamiltonian is given by

\begin{equation}
\label{eq:15}
H_{\text{int}}(t) = \sum_if^{(i)}(t)\sigma^{(i)}
\end{equation}
where $\sigma^{(i)}$ are operators, and the correlations between the $f^{(i)}(t)$ are given by a covariance matrix $R_{ij}(\tau)$
\begin{equation}
\label{eq:16}
R_{ij}(\tau) \equiv \expval{f^{(i)}(t)f^{(j)}(t-\tau)}
\end{equation}
The perturbation Hamiltonian in the Floquet basis is then
\begin{equation}
\label{eq:17}
\Hi = \sum_{\alpha, \beta, k, i}e^{i\Delta_{\alpha \beta k}t}\ket{\Phi_{\alpha}(0)}\bra{\Phi_{\beta}(0)}f^{(i)}(t)X_{\alpha \beta k}^{(i)}
\end{equation}
where
\begin{equation}
  \bra{\Phi_{\alpha}(t)} \sigma^{(i)} \ket{\Phi_{\beta}(t)} = \sum_{k=-\infty}^{\infty}e^{ik\omega t}X_{\alpha \beta k}^{(i)}
\end{equation}

Following the derivation for the single field case, the master equation can now be written as
\begin{multline}
\frac{d\tilde{\rho}}{dt} = \frac{1}{(i\hbar)^2}  \sum_{\alpha, \beta, k, i} \sum_{\alpha', \beta', k', j} J_{ij}(\Delta_{\alpha' \beta' k'}) e^{i(\Delta_{\alpha, \beta, k} - \Delta_{\alpha' \beta' k'})t} X_{\alpha \beta k}^{(i)} X_{\alpha' \beta' k'}^{(j)*} \\
\left[\Gamma_{\alpha\beta}, \left[ \Gamma_{\beta'\alpha'}, \tilde{\rho}(t) \right] \right]
\end{multline}

where
\begin{equation}
\label{eq:18}
J_{ij}(\omega) \equiv \int_{0}^{\infty}R_{ij}(\tau)e^{i\omega\tau}d\tau
\end{equation}

Once again, we specialize to two cases:

\subsubsection{$(\alpha, \beta, k) = (\alpha', \beta', k')$}
Then we can define the matrix
\begin{equation}
\label{eq:19}
\zeta_{\alpha\beta} = \frac{1}{\hbar^{2}}\sum_{k,i,j}J_{ij}(\Delta_{\alpha \beta k})X_{\alpha \beta k}^{(i)} X_{\alpha \beta k}^{(j)*}
\end{equation}
And the rest of the derivation proceeds as in Section~\ref{subsec:rwa}.

\subsubsection{$\alpha = \beta, \alpha' = \beta', k = k', \alpha \neq \alpha'$}
Then we can define the matrix
\begin{equation}
\label{eq:20}
\xi_{\alpha\alpha'} = \frac{1}{\hbar^{2}}\sum_{k,i,j}J_{i,j}(k\omega)X_{\alpha \alpha k}^{(i)} X_{\alpha' \alpha' k}^{(j)*}
\end{equation}
And the rest of the derivation proceeds as in Section~\ref{subsec:rwa2}

\section{Cross-Spectrum Properties}
\begin{equation}
\label{eq:27}
J_{xy}(\omega) \equiv \int_{0}^{\infty}\expval{x(t)y(t + \tau)} e^{-i\omega \tau}d\tau
\end{equation}

\begin{equation}
  \label{eq:29}
  \boxed{J_{xy}(-\omega) = \int_{0}^{\infty}\expval{x(t)y(t + \tau)} e^{+i\omega \tau}d\tau = J_{xy}(\omega)^{*}}
\end{equation}

\begin{align}
  J_{xy}(\omega) + J_{yx}^{*}(\omega) &= \int_{0}^{\infty}\expval{x(t)y(t + \tau)} e^{-i\omega \tau} + \expval{y(t)x(t + \tau)} e^{+i\omega \tau} d\tau \\
                            &= \int_{0}^{\infty}\expval{x(t)y(t + \tau)} e^{-i\omega \tau} d\tau + \int_{-\infty}^0\expval{y(t)x(t - \tau)} e^{-i\omega \tau} d\tau \\
                            &= \int_{-\infty}^{\infty}\expval{x(t)y(t + \tau)}e^{-i\omega \tau} d\tau
\end{align}
\begin{equation}
\boxed{J_{xy}(\omega) + J_{yx}^{*}(\omega) = S_{xy}(\omega)}
\end{equation}

\bibliographystyle{unsrtnat}
\bibliography{references}
\end{document}